\documentclass[pre,twocolumn,superscriptaddress]{revtex4}
\usepackage{epsfig,amsmath}
\begin {document}
\title {Dynamics of the breakdown of granular clusters}
\author{Fran\c{c}ois Coppex}
\affiliation{Department of Physics, University of Gen\`eve, CH 1211
Gen\`eve 4, Switzerland}
\author{Michel Droz}
\affiliation{Department of Physics, University of Gen\`eve, CH 1211
Gen\`eve 4, Switzerland}
\author{Adam Lipowski}
\affiliation{Department of Physics, University of Gen\`eve, CH 1211
Gen\`eve 4, Switzerland}
\affiliation{Department of Physics, A.~Mickiewicz University,
61-614 Pozna\'{n}, Poland}
\begin {abstract}
Recently van der Meer et al. studied the breakdown of a granular 
cluster (Phys.~Rev.~Lett.~{\bf 88}, 174302 (2002)).
We reexamine this problem using an urn model, which takes into account fluctuations and finite-size effects. 
General arguments are given 
for the absence of a continuous transition when the number of urns 
(compartments) is greater than two. 
Monte Carlo simulations show that the lifetime of a cluster $\tau$ 
diverges at the limits of stability as $\tau\sim N^{1/3}$, where $N$ is 
the number of balls. 
After the breakdown, depending on the dynamical rules of our urn model, 
either normal or anomalous diffusion of the cluster takes place.
\end{abstract}
\maketitle


\section{introduction}
Dissipation of kinetic energy during inelastic collisions in gaseous 
granular systems has profound consequences~\cite{SWINNEY,MUZZIO}.
One of the most spectacular ones is formation of spatial 
inhomogeneities~\cite{GOLD}, which drastically contrast with a uniform 
distribution of molecules or atoms whose dynamics is essentially elastic.

Some time ago Schlichting and Nordmeier presented a simple experiment which
demonstrates some consequences of inelasticity of granular systems~
\cite{SCHLICH}.
They used a container separated into two equal compartments by a wall which 
has a narrow horizontal slit at a certain height.
The container is filled with balls (plastic or metallic) and subjected to 
vertical shaking.
For vigorous shaking the balls distribute equally  between two compartments.
However, when the shaking is sufficiently mild, a nonsymmetric distribution
occurs.
In such a case the compartment with majority of balls, due to numerous 
inelastic collisions, is effectively cooler than the other one.
Consequently, less balls are leaving this compartment which stabilizes such
an asymmetric distribution of balls.
To explain this experiment, Eggers derived a phenomenological equation
for the flux $F(n)$ of balls leaving a given compartment~\cite{EGGERS}
\begin{equation}
F(n)=Cn^2{\rm exp}(-Bn^2).
\label{eggers}
\end{equation}
In the above equation $n$ is the concentration of balls in a given urn and $B$ 
and $C$ are 
constants which depend on the properties of balls, typical sizes of the 
system and of parameters of shaking (the constant $C$ may
be eliminated by an appropriate redefinition of the time scale).
In agreement with experiment, eq.~(\ref{eggers}) predicts for sufficiently 
large $B$ unequal distribution  of balls.
The above experiment was repeated in the case when the number of compartments
$L$ was greater than two by van der Meer et al.~\cite{MEER1}.
In such a case formation of unequal distribution of balls is accompanied by 
strong hysteresis which is in agreement with theoretical analysis~\cite{MEER2}.
Moreover, certain aspects of these phenomena for $L=2$ were approached 
using hydrodynamic equations~\cite{BREY}.
  
Recently, van der Meer et al. examined the case of $L>2$ 
further~\cite{MEER3}.
In particular, they studied dynamics of configurations (clusters) starting 
from all balls localized in a single compartment.
Using a  theoretical model based on eq.~(\ref{eggers}), they have 
shown that when shaking is strong enough such a cluster breaks down
and diffuses with the anomalous diffusion exponent 1/3 (in the following we refer to this model as MWL).
For less vigorous shaking, the cluster remains relatively stable and only 
after some time it abruptly breaks down. 
Some of their predictions were confirmed experimentally.

In the framework of the MWL model it is 
rather difficult to include the effect of fluctuations.
Such fluctuations might originate due to for example a finite number of 
balls and especially close to critical points they might play an important role.
In an attempt to take such effects into account a generalization of 
Ehrenfest's~\cite{EHREN} urn model was recently examined in the case 
$L=2$~\cite{LIPDROZ}.
Relative simplicity of the model allows for a detailed study of its various
characteristics.

The motivation of the present paper is to re-examine the breakdown of 
granular clusters using the urn model in the case $L>2$.
In section~\ref{section2} we define the model and present its steady-state 
phase diagram for $L=3$.
We also argue that, in analogy to the Potts model in the mean-field limit,
there are no continuous transitions for $L>2$.
In section~\ref{section3} we examine dynamics of the breakdown of clusters 
in a similar way as van der Meer et al.~\cite{MEER3}.
Although qualitatively our results are similar to theirs, in our model the 
diffusion of the cluster is normal with the exponent 1/2. 
Moreover, we calculate 
the size dependence of the lifetime of a cluster $\tau$ and show that at 
the limits of stability it scales as $N^{1/3}$.
In section~\ref{section4} we present a modified version of the urn model 
which in the steady state reproduces the flux (\ref{eggers}). 
The diffusion of the broken down cluster is then shown to be anomalous 
with exponent $1/3$, as it was already found~\cite{MEER3}.
It was suggested that essential features of the MWL model are independent on 
the precise form of the flux (\ref{eggers}), as long as it has a single 
hump~\cite{MEER3}.
On the contrary, our results show that at least the diffusion exponent 
depends on some details 
of the flux and not only on its qualitative shape (in our models the flux 
is also a single hump function).
Section~\ref{section5} contains our conclusions.

\section{Model and its steady-state properties}\label{section2}
Our model is a straightforward generalization of the two-urn 
case~\cite{LIPDROZ}:
$N$ particles are distributed between $L$ urns and the number of
particles in $i$-th urn is denoted as $N_i$ ($\sum_{i=1}^{L} N_i=N$).
Urns are connected through slits sequentially: $i$-th urn is connected with 
$(i-1)$-th and $(i+1)$-th.
Moreover, periodic boundary conditions are used, i.e., first and 
$L$-th urns are connected.
Particles in a given urn (say $i$-th) are subject to thermal fluctuations and 
the temperature $T$ of this urn depends on the number of particles in it as:
\begin{equation}
T(n_i)=T_0+\Delta(1-n_i),
\label{temp}
\end{equation}
where $n_i$ is a fraction of the total number of particles in a 
given urn ($n_i=N_i/N$) and $T_0$ and $\Delta$ are positive constants. 
Equation (\ref{temp}) is the simplest function which reproduces the fact 
that due to inelastic collisions between particles, their effective 
temperature decreases as their number in a given urn increases.
Next, we define the dynamics of the model as:
\begin{enumerate}
\item[(i)] One of the $N$ particles is selected randomly.
\item[(ii)] With probability ${\rm exp}[-1/T(n_i)]$ the selected 
particle is placed in a randomly chosen neighboring urn, where $i$ is the 
urn of a selected particle.
\end{enumerate}

\begin{figure}
\centerline{\epsfxsize=9cm 
\epsfbox{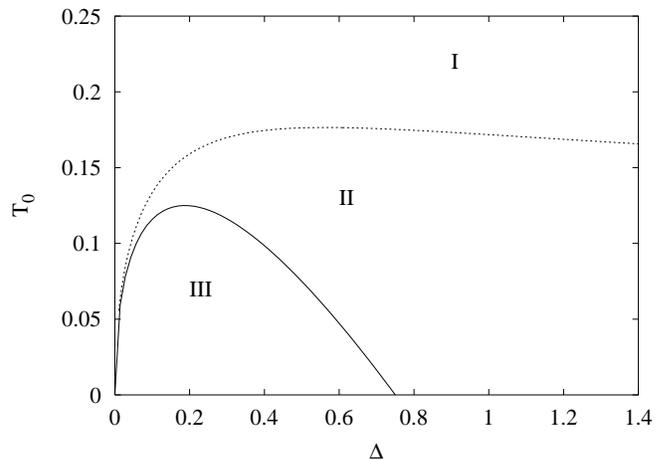}
}
\caption{The steady-state phase diagram for the three-urn model.
See text for a description of phases.}
\label{diagram}
\end{figure}

The above rules implies that the flux of particles leaving $i$-th urn is, 
up to a proportionality constant, given by
\begin{equation}
F(n_i)=n_i \, {\rm exp}\left[-\tfrac{1}{T(n_i)}\right],
\label{flux}
\end{equation}
where $T(n_i)$ is defined in (\ref{temp}).
Let us notice that the flux (\ref{flux}), similarly to (\ref{eggers}), is a 
single hump function.
Having an expression for the flux we can write the equations of motion as:
\begin{equation}
\frac{dn_i}{dt}=\tfrac{1}{2}F(n_{i-1})+\tfrac{1}{2}F(n_{i+1})-F(n_{i}),
\label{motion}
\end{equation}
where $i=1,2,...,L$. 
Steady-state properties of this model can be obtained using similar
analysis as in the $L=2$ case~\cite{LIPDROZ} or as for $L>2$  but with fluxes 
given by eq.~(\ref{eggers})~\cite{MEER2}.
The results of this analysis in the $L=3$ case are presented in 
Fig.~\ref{diagram}.
In region I and II the symmetric phase ($n_1=n_2=n_3=1/3$) is stable.
The continuous line in Fig.~\ref{diagram}, which locates the limit of 
stability of this phase, is given by the following equation:
\begin{equation}
T_0=\sqrt{\frac{\Delta}{3}} -\frac{2\Delta}{3}.
\label{critical}
\end{equation}
This equation has a very similar form to the corresponding equation in the 
$L=2$ case~\cite{LIPDROZ}.
Asymmetric solution, where one of the urns has the majority of balls and 
remaining two urns have only a small equal fraction of balls ($n_1>n_2=n_3$), 
is stable in region II and III.
The line separating regions I and II can be determined only numerically as a
solution of a transcendental equation, similarly to the $L=2$ 
case~\cite{LIPDROZ}.
There is also a third type of solution where two urns contain majority 
of balls and the third urn has only a small fraction of them ($n_1=n_2>n_3$).
Such a solution, which has saddle-like stability, exists only in region III.
Similar solutions can be found for the MWL model~\cite{MEER1,MEER2}.
An important, qualitative difference with the case $L=2$, is that 
regions I and III are always separated by region II where both 
symmetric and asymmetric solutions are stable, hence the 
tricritical point is located at the origin $T_0 = \Delta = 0$. 
It means that a phase transition between these two phases is always 
accompanied by hysteresis effects. 
On the other hand in the $L=2$ case continuous transitions are possible, 
which are not accompanied by hysteresis~\cite{LIPDROZ}. 
Such a behaviour is actually in agreement with experimental data and with MWL 
model~\cite{MEER1}.

Has this qualitative difference a more general explanation or is it 
rather a coincidental property?
In our opinion, absence of continuous transitions for $L>2$ is a generic
property of such systems and at least to some extent could be understood.
First, let us notice that the phase transition for $L=2$ is a manifestation
of the spontaneous symmetry breaking in the system: in certain regime one
of the two identical urns is preferentially filled with balls.
Such a situation resembles the phase transition in the $S=1/2$ Ising model, 
where below certain temperature the up-down symmetry is broken and the system
acquires spontaneous magnetization~\cite{HUANG}.
Actually, this analogy can be confirmed more quantitatively.
We have shown that for $L=2$ and at the critical point the probability 
distributions
has the same moment ratios as in the Ising model in dimension $d$ greater 
than the so-called upper critical dimension ($d>4$)~\cite{LIPDROZ2}. 
Let us notice, that in our model balls are selected randomly which means that
this is essentially a mean-field model.
Moreover, our model is a dynamical, spaceless model, contrary to the Ising 
model, which is a lattice equilibrium model.
The fact that such different models have some similarities shows that as far
as the critical behaviour is concerned what really matters is symmetry.
In both cases this is the $Z_2$ symmetry which is broken below the critical 
point.

Pushing this analogy further, we expect that for $L>2$ the phase transition in 
our model should be similar to the phase transition of the $L$-state Potts 
model above the critical dimension~\cite{WU}.
In the $L$-state Potts model at 
sufficiently low temperature one of the $L$ symmetric ground states is 
preferentially selected. 
However, it is well-known that above the upper critical dimension and for $L>2$ there 
are only discontinuous transitions in the Potts model~\cite{WU}.
Consequently, the transition in the urn model, and most likely in related 
models which preserves $Z_L$ symmetry of compartments, should be discontinuous.

Let us notice that one can easily break the symmetry of the compartments 
e.g., changing the boundary conditions, which in our analogy introduces some 
asymmetry in the Potts model.
It is possible that in such a case the system effectively will become 
similar to the $L=2$ system and will exhibit a continuous transition.
Finally, we expect that for $L>3$ the phase diagram should be topologically similar to the one for $L=3$ shown in Fig.~\ref{diagram}.


\section{Dynamical properties of cluster configurations}\label{section3}
\begin{figure}
\centerline{\epsfxsize=9cm 
\epsfbox{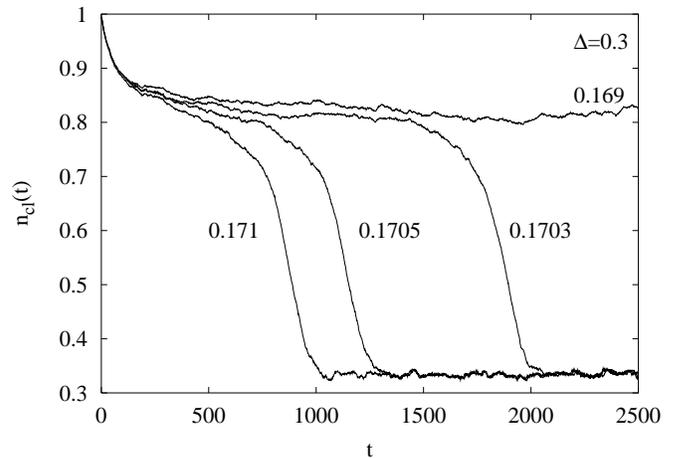}
}
\caption{The time evolution of the fraction of balls of the cluster
$n_{cl}$ close to the limits of stability of the asymmetric phase 
($N=5\cdot 10^4,\ L=3$).
The values of $T_0$ are indicated.
For $\Delta=0.3$ the limit of stability of the asymmetric phase is at 
$T_0=0.169829772...$
For a larger number of balls $N$, stochastic fluctuations will diminish.}
\label{time}
\end{figure}

In the present section we study certain dynamical properties of cluster 
configurations.
We used Monte Carlo simulation.
Since it is rather straightforward, we omit a more detailed description of the
numerical implementation of the dynamical rules of our model.
Initially, we place all balls in one urn and examine its subsequent evolution.
If the parameters $T_0$ and $\Delta$ are such that the system is in region I then
such a cluster is unstable and after some time due to fluctuations it breaks 
down and balls spread throughout all urns.
This is illustrated in Fig.~\ref{time} which shows the concentration of balls in
the urn in which the balls were initially placed.
Let us notice that (i) the breakdown is relatively abrupt and during the 
evolution up to 
the breakdown the concentration of balls only slightly decreases; (ii) upon 
approaching the line separating regions I and II the 
lifetime of the cluster $\tau$ increases.
\begin{figure}
\centerline{\epsfxsize=9cm 
\epsfbox{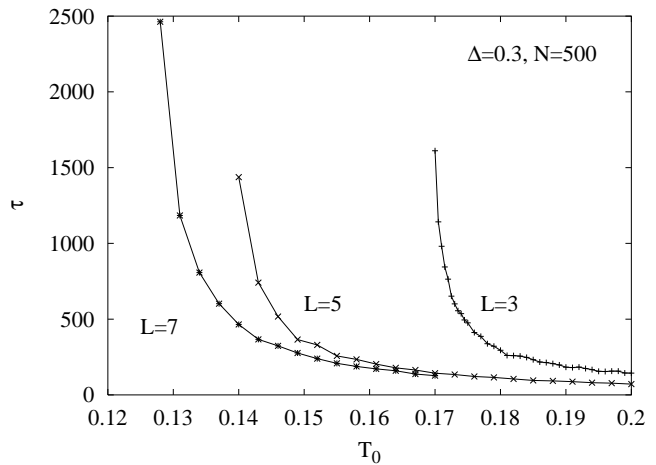}
}
\caption{
The average lifetime of a cluster $\tau$ as a function of $T_0$ for different number of urns $L$. Each point is an average of at least 300 runs.}
\label{taut0}
\end{figure}
Since in region II the asymmetric state has an infinite lifetime it means that
$\tau$ must diverge upon approaching this region.
This behaviour is seen in Fig.~\ref{taut0}.
In addition to the three-urn case we also made analogous measurements of 
$\tau$ for $L=5$ and 7 and the results are also shown in Fig.~\ref{taut0}.
Let us notice that results presented in Fig.~\ref{time} and Fig.~\ref{taut0} 
are similar to those obtained by van der Meer~\cite{MEER3}, although they are parametrized by a different variable.

\begin{figure}
\centerline{\epsfxsize=9cm 
\epsfbox{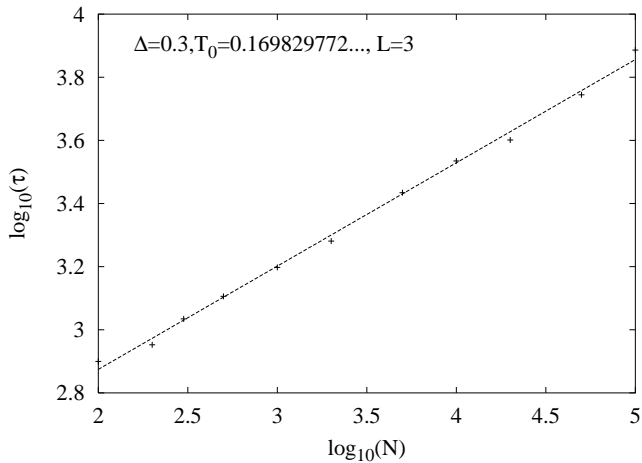}
}
\caption{
The average lifetime of a cluster $\tau$ as a function of the number of balls 
$N$ at the limits of stability of the asymmetric phase.
Each point is an average of at least 300 runs.}
\label{tau}
\end{figure}

The limit of stability of the asymmetric phase can be regarded as a 
critical point.
Thus, we expect that exactly at this point e.g., the lifetime $\tau$ has a 
power-law divergence $\tau=N^z$, and $z>0$.
Such a behaviour is shown in Fig.~\ref{tau}.
From the slope of the straight line, which is a least-square fit to 
our data we estimate $z=0.32(3)$.
Let us notice that in the two-urn model at the limits of stability $\tau$
exhibits a very similar divergence~\cite{LIPDROZ}.
In the case $L=2$ more precise calculations were possible strongly  
suggesting that $z=1/3$ which is also consistent with the present 
three-urn model result. Let us emphasize that because in our model the number of balls is finite, we can study size dependent quantities as shown in Fig.~\ref{tau}. Such calculations would not be possible for models solely based on steady-state equations.

\begin{figure}
\centerline{\epsfxsize=9cm 
\epsfbox{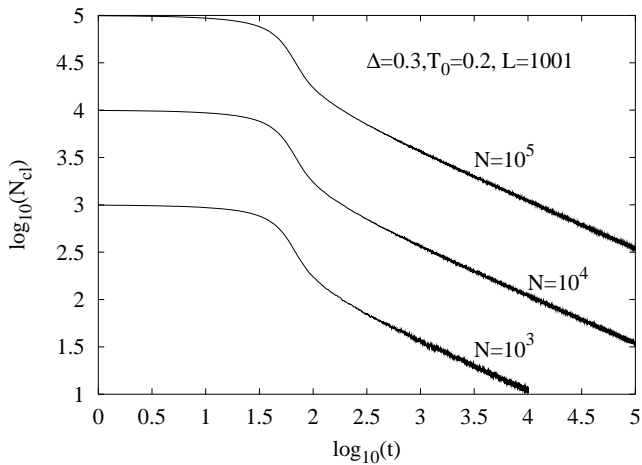}
}
\caption{
The average occupancy of a central urn $N_{cl}$ as a function of time $t$.
The slope of decay is very close to 0.5 which confirms the diffusive nature
of spreading ($N_{cl}\sim t^{-1/2}$).
Each curve is obtained from averaging over 50 independent runs.}
\label{timecentral}
\end{figure}

Finally, let us examine the breakdown of a cluster in the many-urn 
case $L\gg 1$.
In such a case a continuous approach to the MWL model shows that
after breaking down, the cluster diffuses with the anomalous exponent 
$1/3$~\cite{MEER3}.
Results of our simulations are shown in Fig.~\ref{timecentral}.
From these data we conclude that spreading of a cluster occurs with the 
ordinary exponent 1/2 rather than anomalously.
Ordinary diffusion in our model can be also easily explained analytically 
applying basically the same continuous approach as used in~\cite{MEER3}.
In this approach the set of equations of motion~(\ref{motion}) is transformed 
into a partial differential equation.
Then, one immediately realizes that the linear term in front of the 
exponent in eq.~(\ref{flux}) leads to the ordinary diffusion equation.
On the other hand, the anomalous diffusion of MWL model can be traced back to 
the quadratic (in $n$) term in the flux in eq.~(\ref{eggers}).
This quadratic term is related with two-particle collisions~\cite{MEER3}.


\section{The pair model}\label{section4}

One can easily construct urn models for which the expression for the flux will 
have a different form. 
In particular, redefining the effective temperature (\ref{temp}) and drawing  
each time a pair of balls we obtain an urn model with the flux of exactly the 
same form as eq.~(\ref{eggers}). 
This dynamics takes into account some of the two particles 
correlations. 
It allows us to recover some properties of the MWL model and establish 
further results.

The model, which we call a pair model, is similar to the previously 
described one, except that its dynamics is now defined as:
\begin{enumerate}
\item[(i)] Two different balls are selected randomly. \\
\item[(ii)] If and only if the two balls are in the same urn, with 
probability $\exp[-B n_i^2]$ the selected balls are placed in the same 
randomly chosen neighboring urn, where $i$ is the urn of the selected 
particles.
\end{enumerate}

One can easily see that the probability that two randomly selected balls belong to the $i$-th urn is given as $\frac{N_i}{N} \cdot \frac{N_i-1}{N-1}$, which for $N \to \infty$ becomes $n_i^2$. Multiplying $n_i^2$ with the transition probability $\exp[-B n_i^2]$ we 
obtain that the flux in the pair model is proportional to 
eq.~(\ref{eggers}). 
It means that as far as the steady-state properties are concerned, 
the pair model is equivalent to the MWL~\cite{MEER1,MEER2}. 
In particular for $L=2$ one easily obtains the critical value $B=4$ for the 
continuous transition between the symmetric ($B<4$) and asymmetric phase 
($B>4$).
For $L=3$ one obtains two critical points $B_1 = 6.552703411\ldots$ and 
$B_2 = 9$. 
The first one can only be determined numerically. 
Similarly to Fig.~\ref{diagram}, for $B < B_2$ the symmetric solution is 
stable whereas for $B > B_1$ the asymmetric solution 
stable. 
In the interval $B \in [B_1,B_2]$ both symmetric and asymmetric solutions 
are stable, which is the interval showing hysteresis with respect to the 
driving parameter $B$.

Qualitatively the dynamical properties of cluster configurations in the pair model 
are similar to those described in previous section. 
In particular for $L=3$ and $B=B_1$, the average lifetime of a cluster 
$\tau$ as a function of the number of balls $N$ once more shows a 
power-law divergence $\tau = N^z$, with $z=0.31(3)$ suggesting that $z=1/3$. 
It shows a certain universality of this exponent with respect to different
dynamical rules.

\begin{figure}
\centerline{\epsfxsize=9cm 
\epsfbox{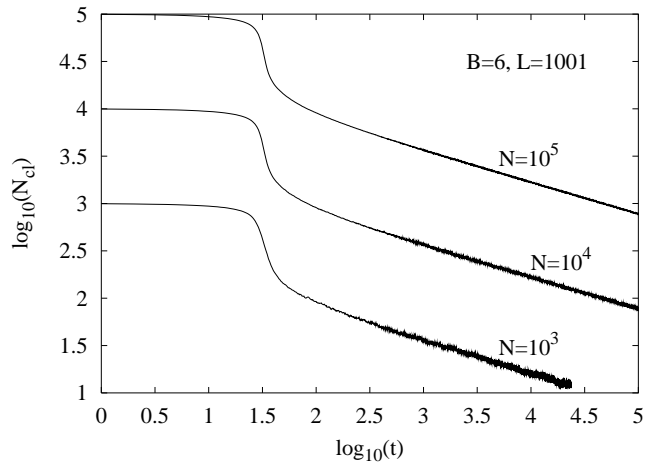}
}
\caption{
The average occupancy of a central urn $N_{cl}$ as a function of 
time $t$ for the pair model.
The slope of decay is very close to $1/3$ which confirms the anomalous 
diffusive nature
of spreading ($N_{cl}\sim t^{-1/3}$). 
Each curve is obtained from averaging over 50 independent runs.}
\label{timecentral2}
\end{figure}

Finally, Fig.~\ref{timecentral2} shows the diffusion of the broken down 
cluster. 
Since the asymptotic slope of our data is very close to 1/3 we conclude
that in this case the diffusion is anomalous, as already predicted by van der 
Meer et al. who used the continuous approach~\cite{MEER3}.

The pair model and the model examined in the previous section exhibit 
qualitatively similar behaviour for most of the physical quantities.
The main difference is the diffusion: it is anomalous in the pair 
model and ordinary in model examined in the previous section.
It would be desirable to experimentally examine the nature of diffusion in
such systems.

\section{Conclusions}\label{section5}
We examined two $L>2$ versions of the $L$-urn model of 
compartmentalization of vibrated sand.
Our models qualitatively recover experimental findings and previous 
steady-state calculations.
In addition, our models take into account fluctuations caused by the 
finite number of balls. 
Using symmetry  properties, we related them with 
high-dimensional Potts model and argued that for $L>2$  phase transitions in
such systems should be discontinuous.
Although several quantities exhibit qualitatively a similar behaviour for the two different versions of the model, there are important differences too.
In particular, these models predict a different diffusion of a 
broken-down  cluster, which could be either ordinary or anomalous.
It shows that the type of diffusion is very sensitive to dynamical rules of the
model, and consequently, to the form of the flux.
\begin{acknowledgments} 
This work was partially supported by the Swiss National Science Foundation
and the project OFES 00-0578 "COSYC OF SENS".
\end{acknowledgments}

\end {document}